\pdfoutput=1
\documentclass{article}
\usepackage{spconf,amsmath,graphicx}

\title{THE MSXF TTS SYSTEM FOR ICASSP 2022 ADD CHALLENGE}
\name{Chunyong Yang, Pengfei Liu, Yanli Chen,  Hongbin Wang,Min Liu}
\address{Mashang Consumer Finance Co., Ltd.}
\begin{document}
%\ninept
%
\maketitle
\begin{abstract}
This paper presents our MSXF TTS system for Task 3.1 of the Audio 
Deep Synthesis Detection (ADD) Challenge 2022. We use an end to end text to speech system, 
and add a constraint loss to the system when training stage. The end to end TTS system is VITS, 
and the pre-training self-supervised model is wav2vec 2.0. 
And we also explore the influence of the speech speed and volume  in spoofing.
The faster speech means  the less the silence part in audio, the easier to fool the detector.
We also find the smaller the volume, the better spoofing ability, though we normalize volume for submission.
Our team is identified as C2, and we got the fourth place in the challenge.
\end{abstract}
\section{Introduction}
\label{sec:intro}

This paper describes the zkj team’s submissions for Task 3.1 of the Audio 
Deep Synthesis Detection (ADD) Challenge 2022\cite{add2022},which is 
an IEEE International Conference on Acoustics, Speech and Signal Processing (ICASSP) 2022 Signal Processing Grand Challenge.
The purpose of the Challenge is to build innovative new technologies 
that can further accelerate and foster research on detecting deep synthesis and manipulated audios. 
The ADD 2022 includes three tracks, the track 3 contains two tasks: track 3.1 Generation task and track 3.2 Detection task. 
We only participant track 3.1, which aims to generate fake audios that can fool the fake detection model.

To generate fake audios, the text to speech or voice cloning system can be used. 
Though there are methods like Adversarial Voice Conversion\cite{ding21_interspeech}, 
it needs to generate  adversarial examples against a neural-network-based spoofing detector by white-box.   
But the task requires that the text content and speaker identity of generated audio is constrained to ensure the submitted sample 
are generated by deep fake methods. So only using Voice Conversion is not suitable for text input required and the spoofing detector is also not known.   
Also considering the speech recognition and speaker recognition technologies would be applied for 
screening each submitted sample, we need the system to provide high MOS score. Finally, we choose to build a multi-speaker text to speech system.

There are many successful multi-speaker text to speech system, like \cite{chen20r_interspeech} \cite{Chen2021AdaSpeechAT} \cite{Ren2020} etc.
Which one to be chosen needs to be determined. 
Consider our purpose is fooling the fake detection model, so we need the distribution of the generated audio is the same as the train data.
The less gap between the fake and real audio the better ability to fool the detector. So we need a GAN vocoder like hifi-gan \cite{Kong2020} which generates waveform like the train data. 
And then we use The full end to end text to speech system to prevent the two stages system producing gap between acoustic model and vocoder, 
Finally, we choose conditional variational autoencoder with adversarial learning for end to end text to speech(VITS) \cite{Kim2021}, 
which gets MOS score almost the same as ground truth.
We are also inspired by the method \cite{Cai2020} that uses speaker verification as a constraint to improve white-box spoofing attack. 
But we replace speaker verification with the wav2vec 2.0 \cite{Baevski2020} for black-box spoofing. 
For using the wav2vec 2.0 which is a self-supervised learning of speech representation, we follow the research \cite{Shah2021}\cite{Choi2021}.
It shows which layer of wav2vec 2.0 is better for a specific category of feature to choose for feature extraction for downstream tasks.

The main contributions of this paper are summarized as follows:

(1)we use wav2vec 2.0 to improve speaker identity of the VITS, it shows better result than origin VITS system.  

(2)we demonstrate the VITS system is better than fastspeech2+hifi-gan system for fooling the detector such as \cite{Tak2021}. And found the suitable model settings for VITS.  

(3)we find that the talking speed of speech can influence the detector EER on test data. We also find that the volume of the audio can influence the result, but we follow the Challenge guide to set it as the original train data's volume. 

This paper is organized as follows: Section 2 describes our system. 
Section 3 introduces the setup of the datasets for the challenge, experiments.
Section 4 gives the result and analysis. 
Finally, Section 5 gives the conclusions.

\section{SYSTEM DESCRIPTION}
\label{sec:system}

Our system is made up of two modules:  text to phoneme front-end module, end to end phoneme to speech module.

\subsection{Text to Phoneme Front-end Module}
\label{ssec:frontend}

Though  texts  and  pinyin  are  both  provided  in  the  training corpus, we just need to convert the challenge texts to phoneme for synthesis.
We do text normalization (TN) first to convert the number and symbols to right pinyin. Here we use rule based method to do TN. 
Then we use g2pM\cite{Park2020g2pMAN} to  process polyphonic characters. 
And finally the rest Chinese characters in texts are  converted to pinyin with 5 tones by a grapheme-to-phoneme (G2P) toolkit named pypinyin.
Our system don't predict prosodic structure because we believe multi-speaker system with enough training data can learn it well, There is also
another reason that some data of the AISHELL-3 don't read prosodic structure rightly. 

\subsection{End to End Phoneme to Speech Module}
\label{ssec:endtoend}
The end to end phoneme to speech module, we use VITS, which's input is phoneme width tones and output audio waveform. 
we keep the VITS model settings in the paper\cite{Kim2021} and change the sample rate to 16K Hz.

To improve the spoofing ability, when training the VITS model, we add a constraint loss to its generate loss. 
The constraint loss is a MSE loss of the fake and real features that are extracted by wav2vec 2.0. As the papers\cite{Shah2021}\cite{Choi2021} give the experiments that the first layer of the wav2vec 2.0 got good result. 
So when train stage, use the first layer of wav2vec 2.0 to extract features of fake and real audio for constraint loss calculating.
The wav2vec 2.0 are trained using the open source data WenetSpeech\cite{ws2022}, a 10000+ hours multi-domain mandarin corpus.

\section{Experiments}
\label{sec:experiments}

\subsection{Datasets}
\label{ssec:datasets}
For the audio fake game generation task, participants are allowed to generate deepfake audio based on the AIShell-3 dataset. 
The AIShell-3 corpus contains roughly 85 hours of speech recordings spoken by 218 native Mandarin Chinese speakers. 
Speech is recorded by a high-quality microphone in a typical room with room reverberation and some small background noise. 
We use the sox tools to down the sample rate to 16K Hz. We use one data for each speaker for validation, and all rest data for train.
10 speakers ID from AIShell-3 dataset are listed as the evaluation speaker ID, and 500 lines text for each speaker synthesis.

And for evaluation, we use RawGAT-ST \cite{Tak2021} for detecting. The RawGAT-ST use the ASVspoof 2017 eval data for training.

\subsection{Model Settings}
\label{ssec:model}
All models used in experiments are adjusted for 16K Hz sample rate. And keeps other parameters the same as original papers.
And we train a fastspeech2+hifi-gan two stages model for comparing with VITS. Here we don't use MFA to generate the alignments which the fastspeech2 model needed.
We use the alignment learning framework in paper\cite{Badlani2021} instead of MFA, which improves alignment convergence speed of existing attention-based mechanisms, simplifies the training pipeline, and makes the models more robust to errors on long utterances.

\section{Results}
\label{sec:result}

We first test VITS and fastspeech2+hifi-gan with 2000 samples from 100 speakers in AISHELL-3 before the organizers give the evaluation speakers.
From the test, We find volume has huge influence on EER. 
So we only show the result after normalize the volume, the VITS got 33.722 EER on RawGAT-ST while the fastspeech2+hifi-gan got 17.342 EER. 
It demonstrates the end to end VITS system more efficient on spoofing than the fs2+hifi-gan two stages system.

After got the evaluation speakers and texts, we do other experiments on the evaluation test, the result show in Table 1. 
We can see the same VITS parameters, only add wav2vec2.0 feature loss will let EER improve more than one point. 
And the VITS has a stochastic duration predictor, 
the noise scale parameter means a scale factor which multiplied to the standard deviation of the prior distribution.
The noise scale d means the standard deviation of input noise of the stochastic duration predictor.   
The two parameters influence the phoneme duration, the smaller the scale, the shorter the phoneme's duration.
We can see that shorter duration means speech faster, and have less silence part, the fake ability will improve.
But when parameter set to zero, the EER will go down, and too fast will influence the MOS, so finally we use the parameters that ns=0.1, nsd=0.3 which the EER is 34.810.

\begin{table}[h!]
  \centering
   \begin{tabular}{||c c||} 
   \hline
   models & EER(RawGAT-ST) \\ [0.5ex] 
   \hline\hline
   VITS(ns=0.667,nsd=0.8) & 30.681 \\ 
   VITS(ns=0.667,nsd=0.8)+wav2vec & 31.819 \\
   VITS(ns=1.0,nsd=0.3)+wav2vec & 20.000 \\
   VITS(ns=0.1,nsd=0.3)+wav2vec & {\bf34.810} \\
   VITS(ns=0.0,nsd=0.3)+wav2vec & {33.000} \\
   VITS(ns=0.1,nsd=1.0)+wav2vec & 34.030 \\ [1ex] 
   \hline
   \end{tabular}
   \caption{Tests on the 10 evaluation speakers and texts. ns means the noise scale factor, nsd means noise standard deviation of the duration stochastic duration predictor. wav2vec means using wav2vec2.0 for VITS generate loss.}
\end{table}

\section{Conclusions}
\label{sec:conclusions}
In this work, we build an end to end multi-speaker text to speech system. 
We use wav2vec2.0 as a constraint for VITS generate loss, which improve the ability of spoofing.
Then we find the speed of speech influence the detector too, the less the silence part in audio, the easier to fool the detector.
We also find volume has a huge influence in spoofing, the smaller the volume, the better spoofing ability, though we normalize volume for submission.
All the work will make us to create a better detector for voice anti-spoofing in the future.

\section{Acknowledgements}
\label{sec:acknowledgements}

This work is supported by Mashang Consumer Finance Co., Ltd.

%\vfill\pagebreak

% References should be produced using the bibtex program from suitable
% BiBTeX files (here: strings, refs, manuals). The IEEEbib.bst bibliography
% style file from IEEE produces unsorted bibliography list.
% -------------------------------------------------------------------------
\bibliographystyle{IEEEbib}
%\bibliography{refs}
\bibliography{reference}

\end{document}